\documentstyle[12pt]{article}
\if@twoside
 \else
 \fi

 \parskip \medskipamount
\newcommand{\be}{\begin{eqnarray}}
\newcommand{\dlq}{\lq\lq}
\newcommand{\ee}{\end{eqnarray}}
\begin{document}


\title{\boldmath The initial energy density of
 gluons produced in very high energy nuclear collisions}

\author{Alex Krasnitz\\
       {\small\it UCEH, Universidade do Algarve,
        Campus de Gambelas, P-8000 Faro, Portugal.}\\ 
        Raju Venugopalan\\ 
        {\small\it Physics Department,
        Brookhaven National Laboratory,
        Upton, NY 11973, USA.}\\
        }
\maketitle
\begin{abstract}
In very high energy nuclear collisions, the 
initial energy of produced gluons per unit area per unit rapidity, 
$dE/L^2/d\eta$, is equal to $f(g^2\mu L)\,(g^2\mu)^3/g^2$, where
$\mu^2$ is proportional to the gluon density per unit area of the
colliding nuclei.  For an SU(2) gauge theory, we perform a
non--perturbative numerical computation of the function $f(g^2\mu L)$. 
It decreases rapidly for small $g^2\mu L$ but varies only by $\sim
25$\%, from $0.208\pm 0.004$ to $0.257\pm 0.005$, for a wide range 
$35.36$--$296.98$ in $g^2\mu L$, including the range 
relevant for collisions at RHIC and LHC. Extrapolating to SU(3), 
we estimate the initial energy per unit rapidity for $Au$--$Au$ collisions 
in the central region at RHIC and LHC.

\end{abstract}

\vspace*{0.2cm} 

By the end of 1999, the Relativistic Heavy Ion collider (RHIC) at BNL
will begin colliding beams of $Au$--ions at $\sqrt{s}=200$
GeV/nucleon. Some years later, the Large Hadron collider (LHC) at CERN
will collide heavy ions at $\sqrt{s}\approx 5.5$ TeV/nucleon. The objective
of these experiments is to understand the properties of very hot and
dense partonic matter in QCD. It is of considerable interest to
determine whether this hot and dense matter equilibrates to briefly
form a plasma of quarks and gluons (QGP)~\cite{QMProc}.

The dynamical evolution of such a system clearly
depends on the initial conditions, namely, the parton distributions in
the nuclei prior to the collision. For partons with transverse momenta
$p_t \gg \Lambda_{QCD}$, cross--sections  
in the standard perturbative QCD approach
may be computed by convolving the parton distributions
of the two nuclei with the elementary parton--parton scattering cross
sections. At the high energies of the RHIC (LHC) collider, hundreds
(thousands) of mini--jets with $p_t$'s of the order of several GeV 
may be formed~\cite{KajMuell}. Final state interactions of
these mini--jets are often described in multiple scattering (see
Ref.~\cite{Wang} and references therein) or in classical cascade
approaches~ (see Ref.~\cite{GeigerZhang} and references therein).
Estimates for the initial energy density in a self screened parton cascade 
approach can be found in Ref.~\cite{KBX}.

At central rapidities, where $x\ll 1$, and $p_t\gg \Lambda_{QCD}$ with
$x$ defined to be $p_t/\sqrt{s}$, parton distributions grow rapidly,
and may even saturate for large nuclei for $x$'s in the range
$10^{-2}$ to $10^{-3}$ relevant for nuclear collisions at RHIC and LHC
respectively~\cite{Muell1,Raj99}. Coherence effects are important
here, and are only included heuristically in the above mentioned
perturbative approaches.

In this letter, we will describe results from a classical effective
 field theory (EFT) approach which includes coherent effects in the
 small $x$ parton distributions of large nuclei~\cite{RajLar}. If the
 parton density in the colliding nuclei is large at small $x$,
 classical methods are applicable. It has been shown recently that a
 renormalization group improved generalization of this effective
 action reproduces several key results in small $x$ QCD: the leading
 $\alpha_S \log(1/x)$ BFKL equation, the double log GLR equation and
 its extensions, and the small $x$ DGLAP equation for quark
 distributions~\cite{JKMW,JKMWV}. It has also been argued, from other
 considerations, that the main results of this model should be general
 results in small $x$ QCD~\cite{Muell1}.

The EFT contains one dimensionful
parameter $\mu^2$, which is the variance of a Gaussian weight over the color
charges $\rho^\pm$ of partons, of each nucleus, 
at rapidities higher than the rapidity of interest.
For central impact parameters, it is determined to be~\cite{GyulassyMclerran}
\be
\mu^2 = {A^{1/3}\over {\pi r_0^2}} \int_{x_0}^1 dx \left({1\over 2 N_c} 
\,\,q(x,Q^2) + {N_c\over {N_c^2-1}}\,\, g(x,Q^2)\right) \,\, ,
\label{colordensity}
\ee
where $xq(x,Q^2)$ and $xg(x,Q^2)$ stand for the {\it nucleon} quark
and gluon structure functions at the resolution scale $Q$ of the
physical process of interest. Also, one has $x_0 = Q/\sqrt{s}$,
$r_0 = 1.12$ fm, and $N_c$ is the number of colors.  From the HERA
data for $q$ \& $g$, one obtains $\mu\leq 1$ GeV for LHC energies and
$\mu \leq 0.5$ GeV at RHIC~\cite{GyulassyMclerran}. The classical
gauge fields, and hence the classical parton distributions, can be
determined analytically~\cite{JKMW,Kovchegov}.  On this basis, it has
been argued recently that the typical transverse momenta scale $Q_s$ in this
model is further in the weak coupling regime, with $Q_s\sim 1$ GeV for
RHIC and $Q_s\sim 2$--$3$ GeV at LHC~\cite{comment0}.

Kovner, McLerran and
Weigert~\cite{KLW} applied the effective action approach to nuclear
collisions. (For an interesting alternative approach, see 
Ref.~\cite{Balitsky}.) Assuming boost invariance, and matching the
equations of motion in the forward and backward light cone, they obtained the
following initial conditions for the gauge fields in the $A^\tau =0$ gauge:
$A_\perp^i|_{\tau=0}=A_1^i+A_2^i$\,, and 
$A^{\pm}|_{\tau=0}={\pm}\,{ig\over 2}\,
x^{\pm}[A_1^i,A_2^i]$. Here $A_{1,2}^i(\rho^\pm)$ ($i=1,2$) are the pure gauge 
transverse gauge fields
corresponding to small $x$ modes of incoming nuclei (with light cone sources
$\rho^\pm\,\delta(x^\mp)$) in the $\theta(\pm x^-) \theta(\mp x^+)$
regions respectively of the light cone.

The sum of two pure gauges in QCD is not a pure gauge--the 
initial conditions therefore give rise to classical gluon radiation in
the forward light cone. For $p_t>>\alpha_S\mu$, the Yang--Mills
equations may be solved perturbatively to quadratic order in
$\alpha_S\mu/p_t$.  After averaging over the Gaussian random sources
of color charge $\rho^\pm$ on the light cone, the perturbative energy
and number distributions of physical gluons were computed by several
authors~\cite{KLW,GyulassyMclerran,DirkYuri,SerBerDir,Guo}.  In the small
$x$ limit, it was shown that the classical Yang--Mills result agreed
with the quantum Bremsstrahlung result of Gunion and
Bertsch~\cite{GunionBertsch}.

In Ref.~\cite{AlexRaj1}, we suggested a lattice discretization of the 
classical EFT, suitable for a non--perturbative numerical solution.
Assuming boost
invariance, we showed that in $A^\tau=0$ gauge, the real time
evolution of the small $x$ gauge fields $A_\perp (x_t,\tau), A^\eta
(x_t,\tau)$ is described by the Kogut--Susskind Hamiltonian in
2+1--dimensions coupled to an adjoint scalar field. The lattice
equations of motion for the fields are then determined
straightforwardly by computing the Poisson brackets. The initial
conditions for the evolution are provided by the lattice analogue of
the continuum relations discussed earlier in the text. We impose 
periodic boundary conditions on an $N\times N$ transverse lattice, where 
$N$ denotes the number of sites. 
The physical linear size of the system is $L=a\,N$, where $a$ is the
lattice spacing. It
was shown in Ref.~\cite{AlexRaj2} that numerical computations on a
transverse lattice agreed with lattice perturbation theory at large
transverse momentum.  For details of the numerical procedure, and
other details, we refer the reader to Ref.~\cite{AlexRaj2}.

In this letter, we will focus on computing the energy density $\varepsilon$ 
as a function of the proper time $\tau$. This computation on the lattice is 
straightforward. Our main result is contained in 
Eq.~\ref{energydensity}. To obtain this result, we compute the 
Hamiltonian density on the lattice for each $\rho^\pm$, and then 
take the Gaussian average (with the weight $\mu^2$) 
over between $40$ $\rho$ trajectories for the larger 
lattices and $160$ $\rho$ trajectories for the smallest ones.

In our numerical simulations, all the relevant physical information is
compressed in $g^2\mu$ and $L$, and in their dimensionless product
$g^2\mu L$~\cite{RajGavai}.  The strong coupling constant $g$ depends
on the hard scale of interest; from Eq.~\ref{colordensity}, we see
that $\mu$ depends on the nuclear size, the center of mass energy, and
the hard scale of interest; $L^2$ is the transverse area of the
nucleus~\cite{comment1}. Assuming $g=2$ (or $\alpha_S=1/\pi$), $\mu
=0.5$ GeV ($1.0$ GeV) for RHIC (LHC), and $L=11.6$ fm for
$Au$--nuclei, we find $g^2\mu L\approx 120$ for RHIC and $\approx 240$
for LHC. (The latter number would be smaller for a smaller value of
$g$ at the typical LHC momentum scale.)  As will be discussed later,
these values of $g^2\mu L$ correspond to a region in which one expects
large non--perturbative contributions from a sum to all orders in
$\sim 6\,\alpha_S\mu/p_t$, even if $\alpha_S\ll 1$.  We should mention
here that deviations from lattice perturbation theory, as a function
of increasing $g^2\mu L$, were observed in our earlier
work~\cite{AlexRaj2}.

We  shall now discuss results from our numerical simulations. 
In Fig.~1, we plot $\varepsilon\tau/(g^2\mu)^3$,
as a function of $g^2\mu\tau$, in dimensionless units, for the
smallest, largest, and an intermediate value in the range of $g^2\mu
L$'s studied. The quantity $\varepsilon\tau$ has the physical
interpretation of the energy density of produced gluons $dE/L^2/d\eta$
only at late times--when $\tau\sim t$. Though $\varepsilon \tau$ goes
to a constant in all three cases, the approach to the asymptotic value
is different. For the smallest $g^2\mu L$, $\varepsilon \tau$ increases 
continously before saturating at late times. 
For larger values of $g^2\mu L$, $\varepsilon\tau$
increases rapidly, develops a transient peak at $\tau\sim 1/g^2\mu$,
and decays exponentially there onwards, satisfying the relation
$\alpha + \beta\,e^{-\gamma\tau}$, to a constant value $\alpha$ (equal
to the lattice $dE/L^2/d\eta$!). The lines shown in the figure are
from an exponential fit including all the points past the peak.  This
behavior is satisfied for all $g^2\mu L \ge 8.84$, independently of $N$.

Given the excellent exponential fit, one can interpret the decay time
$\tau_D=1/\gamma/g^2\mu$ 
as the appropriate scale controlling the formation of gluons
with a physically well defined energy. In other words, $\tau_D$ is the
``formation time''in the sense used by Bjorken~\cite{Bj,comment2}.  
In Table~1, we tabulate $\gamma$ versus $g^2\mu
L$ for the largest $N\times N$ lattices~\cite{comment3} for all but the 
smallest $g^2\mu L$. For large
$g^2\mu L$, the formation time decreases with increasing $g^2\mu L$,
as we expect it should.
 
In Fig.~2, we plot the asymptotic values $\alpha$ of $\varepsilon
\tau/(g^2\mu)^3$ as a function of $g^2\mu a$ for various values of
$g^2\mu L$. As shown in the upper part of Fig.~2, for smaller $g^2\mu
L$, one can go very close to the continuum limit with excellent
statistics (over $160$ independent $\rho$ trajectories for the two
smallest values of $g^2\mu L$).  In the lower part of Fig.~2, all the
data give straight line fits with good $\chi$--squared values.
We use these fits to extrapolate the value of $\alpha$ in the continuum 
limit. We note that the largest value of $g^2\mu L$ with
the smallest $g^2\mu a$ equal to $0.247$, is relatively much further
away from the continuum limit than the points in the upper part of the
figure.  It is obtained by averaging $40$ independent trajectories on
a $1200\times 1200$ lattice. To lower $g^2\mu a$ below $0.1$, would
require going to lattices with $3000\times 3000$ sites. This exceeds
the CPU memory of our current computational resources. Nevertheless,
even for the largest $g^2\mu L$, we do get a fine linear fit--though we
would warn of a potentially large systematic error in the extrapolated
value of $\varepsilon \tau/(g^2\mu)^3$.

The physical energy per unit area per unit rapidity of 
produced gluons can be defined in terms of a function $f(g^2\mu L)$ as
\be
{1\over L^2}\, {dE\over d\eta} = {1\over g^2}\,f(g^2\mu L)\,(g^2\mu)^3 \, .
\label{energydensity}
\ee
The function $f$ here is obtained by extrapolating the values in
Fig.~2 to the continuum limit.  In Fig.~3, we plot the striking
behavior of $f$ with $g^2\mu L$. For very small $g^2\mu L$'s, it
changes very slightly but then changes rapidly by a factor of two from
$0.427$ to $0.208$ when $g^2\mu L$ is changed from $8.84$ to
$35.36$. From $35.36$ to $296.98$, nearly an order of magnitude in
$g^2\mu L$, it changes by $\sim 25$\%. The precise values of $f$ and
the errors are tabulated in Table~1.
\begin{table}[h]
\centerline{\begin{tabular}{|lrrrrr|} \hline
$g^2\mu L$ & 5.66 & 8.84 & 17.68 & 35.36 & 70.7 \\
$f$ & $.436\pm .007$ & $.427\pm .004$ & $.323\pm .004$ & $.208\pm .004$ 
& $.200\pm .005$ \\
$\gamma$ &  & $.101\pm .024$ & $.232\pm .046$ & $.165\pm .013$ & $.275\pm 
.011$ \\
\hline
$g^2\mu L$ & 106.06 & 148.49 & 212.13 & 296.98 & \\
$f$ & $.211\pm .001$ & $.232\pm .001$ & $.234\pm .002$ & $.257\pm .005$ & \\ 
$\gamma$ & $.322\pm .012$ & $.362\pm .023$ & $.375\pm .038$ & $.378\pm .053$ 
& \\
\hline
\end{tabular}}
\caption{The function $f=dE/L^2/d\eta$ and the relaxation rate 
$\gamma=1/\tau_D/g^2\mu$ 
tabulated as a function of $g^2\mu L$. $\gamma$ has no entry for the 
smallest $g^2\mu L$ since there
$\varepsilon\tau/(g^2\mu)^3$ 
vs $g^2\mu\tau$ differs qualitatively from the other $g^2\mu L$ 
values.}
\end{table}

What is responsible for the dramatic change in the behavior of $f$ as
a function of $g^2\mu L$?  In $A^\tau=0$ gauge, the dynamical
evolution of the gauge fields depends entirely on the initial
conditions, namely, the parton distributions in the wavefunctions of
the incoming nuclei~\cite{KovMueller}.  In the nuclear wavefunction,
at small x, non--perturbative, albeit weak coupling, effects become
important for transverse momenta $Q_s\sim 6\,\alpha_s\mu$. The EFT
predicts that classical parton distributions which have the
characteristic Weizs\"acker--Williams $1/p_t^2$ behaviour for large
transverse momenta ($p_t\gg Q_s$) grow only logarithmically for
$p_t\leq Q_s$. One can therefore think of $Q_s$ as a saturation
scale~\cite{comment0} that tempers the growth of parton distributions
at small momenta.

Now on the lattice, $p_t$ is defined to be $2\pi
n/L$, where $n$ labels the momentum mode.  The condition that momenta
in the wavefunctions of the incoming nuclei have saturated, $p_t\sim
6\,\alpha_S\mu$, translates roughly into the requirement that $g^2\mu L
\geq 13$ for $n=1$.  Thus for $g^2\mu L= 13$, one is only beginning to
sample those modes.  Indeed, this is the region in $g^2\mu L$ in which
one sees the rapid change in $f$.  The rapid decrease in $f$ is likely
because the first non--perturbative corrections are large, and have a
negative sign relative to the leading term. Understanding the later
slow rise and apparent saturation with $g^2\mu L$ requires a better
understanding of the number and energy distributions with $p_t$. This
work is in progress and will be reported on
separately~\cite{AlexRaj3}.

Our results are consistent with an estimate by
A. H. Mueller~\cite{Muell2} for the number of produced gluons per unit
area per unit rapidity. He obtains $dN/L^2/d\eta =
c\,(N_c^2-1)\,Q_s^2/4\pi^2 \,\alpha_S\,N_c$, and argues that the
number $c$ is a non--perturbative constant of order unity. If most of
the gluons have $p_t\sim Q_s$, then $dE/L^2/d\eta =
c^\prime\,(N_c^2-1)\,Q_s^3/4\pi^2\,\alpha_S\,N_c$ which is of the
same form as our Eq.~\ref{energydensity}.  In the $g^2\mu L$ region of
interest, our function $f\approx 0.23$--$0.26$. 
Using the relation between $Q_s$
and $g^2\mu$~\cite{comment0}, we obtain $c^\prime = 4.3$--$4.9$. 
Since one expects a distribution in momenta about $Q_s$, it is very likely that
$c^\prime$ is at least a factor of $2$ greater than $c$--thereby
yielding a number of order unity for $c$ as estimated by Mueller. This
coefficient can be determined more precisely when we compute the
non--perturbative number and energy distributions.

We will now estimate the initial energy per unit rapidity of produced
gluons at RHIC and LHC energies. We do so by extrapolating from our
SU(2) results to SU(3) assuming the $N_c$ dependence to be
$(N_c^2-1)/N_c$ as in Mueller's formula. At late times, the energy
density is $\varepsilon = (g^2\mu)^4\,f(g^2\mu L)\,\gamma(g^2\mu
L)/g^2$, where the formation time is $\tau_D=1/\gamma(g^2\mu
L)/g^2\mu$ as discussed earlier. We find that $\varepsilon^{RHIC}\approx
66.49$ GeV/fm$^3$ and $\varepsilon^{LHC}\approx 1315.56$
GeV/fm$^3$. Multiplying these numbers by the initial volumes at the
formation time $\tau_D$, we obtain the classical Yang--Mills estimate
for the initial energies per unit rapidity $E_T$ to be $E_T^{RHIC}\approx
2703$ GeV and $E_T^{LHC}\approx 24572$ GeV respectively.
  
Compare these numbers to results presented recently by
Kajantie~\cite{Keijo} for the mini--jet energy (computed for 
$p_t > p_{sat}$, where $p_{sat}$ is a saturation scale akin to $Q_s$). He
obtains $E_T^{RHIC} = 2500$ GeV and $E_T^{LHC}=12000$. The remarkable
closeness between our results for RHIC is very likely a
coincidence. Kajantie's result includes a $K$ factor of $1.5$--estimates 
range from $1.5$--$2.5$~\cite{KeijoKari}. If we pick a recent value of 
$K\approx 2$~\cite{Andrei}, we obtain as 
our final estimate, $E_T^{RHIC}\approx 5406$ GeV and $E_T^{LHC}\approx 
49144$ GeV.

To summarize, we performed a non--perturbative, numerical computation,
for a SU(2) gauge theory, of the initial energy, per unit rapidity, of
gluons produced in very high energy nuclear collisions.  Extrapolating
our results to SU(3), we estimated the initial energy per unit
rapidity at RHIC and LHC. We plan to improve our estimates by
performing our numerical analysis for SU(3). Moreover, computations in
progress to determine the energy and number distributions should
enable us to match our results at large transverse momenta to
mini--jet calculations~\cite{AlexRaj3}.
 
\section*{Acknowledgments}

We would like to thank Dr. Frank Paige and Dr. Efstratios Efstathiadis
for their help in using the BNL CCD Linux cluster. R.V. acknowledges
the support of DOE Nuclear Theory at BNL and A.K. acknowledges the
support of the Portuguese Funda\c c\~ao para a Ci\^encia e a
Technologia, grant CERN/P/FIS/1203/98. A.K. is grateful to the BNL
Physics Department for hospitality during the course of this work.
R.V. would like to thank Larry McLerran and Al Mueller for useful
discussions.


\begin{thebibliography}{10}
\bibitem{comment1}These considerations are also valid for non--central 
collisions of identical nuclei. For nuclei with unequal $A$, one has 
$\mu_1$ and $\mu_2$ for the two nuclei. In this case, perturbatively one 
expects $\mu = \sqrt{\mu_1\,\mu_2}$.
\bibitem{comment2}In general, $\varepsilon \tau$ has the physical 
interpretation of an energy only at asymptotic $\tau$. We see that 
it converges to a constant value at a finite $\tau$--and that the rate of 
convergence is controlled by $\tau_D$.
\bibitem{comment3} The central values of $\gamma$ are from uncorrelated fits
to the $g^2\mu\tau$ history of $\varepsilon\tau/(g^2\mu)^3$. 
The errors on $\gamma$ were obtained 
by multiplying the naive errors from the uncorrelated fit by the square root 
of the number of degrees of freedom. This procedure suffices for a qualitative
estimate of the gluon energy density in this work. A more rigorous error 
analysis would require a correlated fit which cannot be reliably performed 
with our current relatively small data samples.
\end{thebibliography}

\newpage
\begin{figure}[ht]
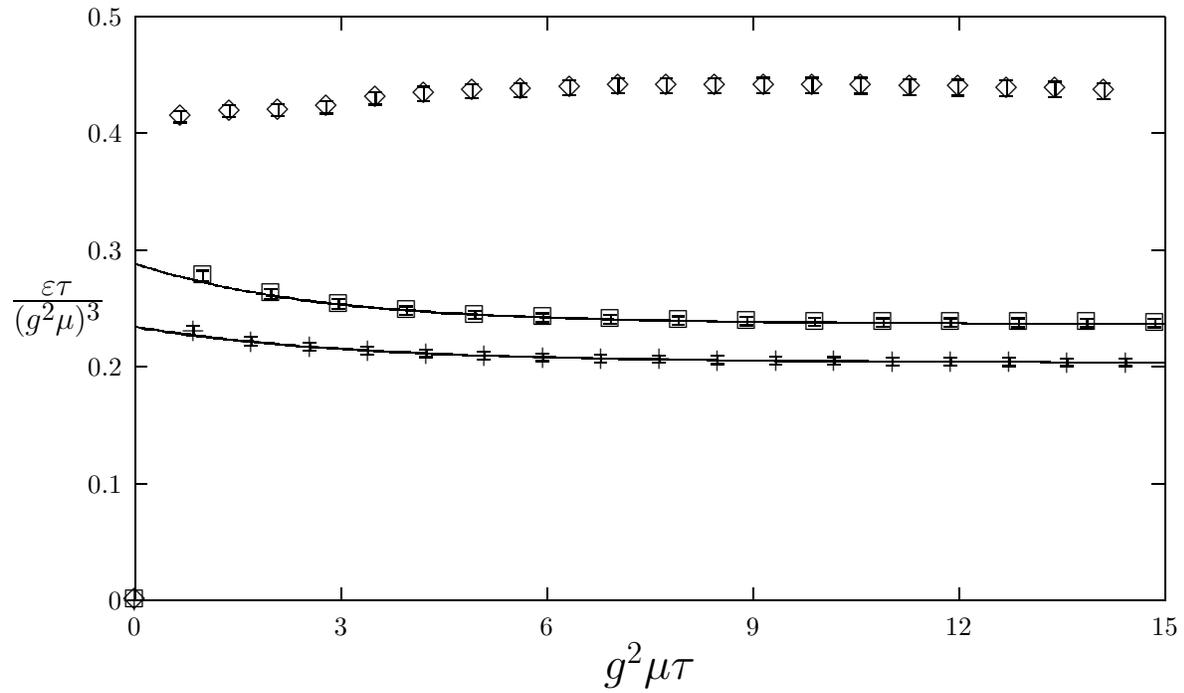

\setlength{\unitlength}{0.240900pt}
\ifx\plotpoint\undefined\newsavebox{\plotpoint}\fi
\sbox{\plotpoint}{\rule[-0.200pt]{0.400pt}{0.400pt}}%

\caption{$\varepsilon\tau/(g^2\mu)^3$ as a function of $g^2\mu\tau$
for $g^2\mu L = 5.66$ (diamonds), $35.36$ (pluses)
and $296.98$ (squares). Both axes are in dimensionless units.
Note that $\varepsilon\tau =0$ at $\tau=0$ for all $g^2\mu L$. The lines are
exponential fits $\alpha + \beta\,e^{-\gamma\tau}$ including all
points beyond the peak.}
\label{eXtR}
\end{figure}

\begin{figure}[hb]
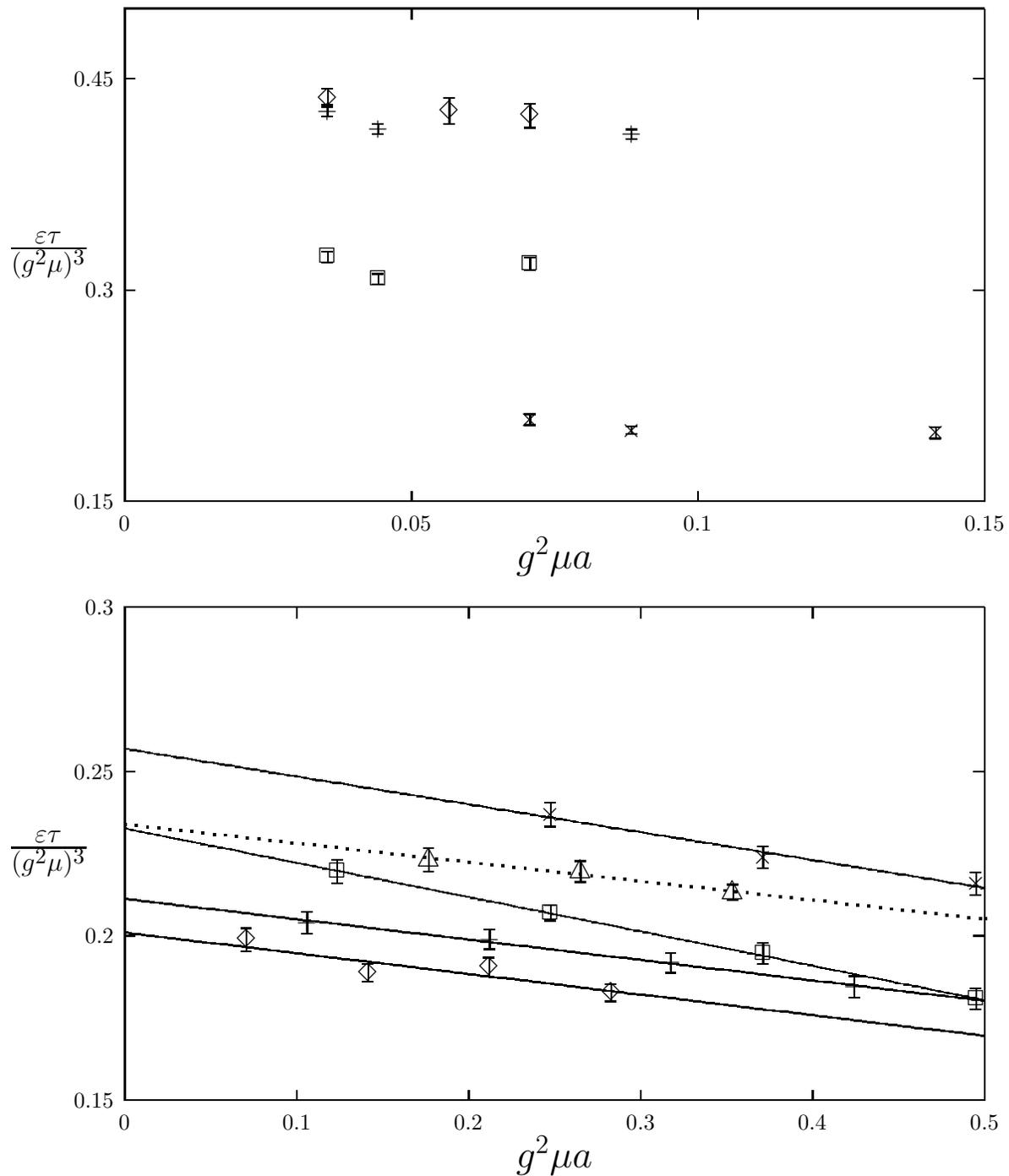

\setlength{\unitlength}{0.240900pt}
\ifx\plotpoint\undefined\newsavebox{\plotpoint}\fi
\sbox{\plotpoint}{\rule[-0.200pt]{0.400pt}{0.400pt}}%

\caption{$\varepsilon\tau/(g^2\mu)^3$ as a function of $g^2\mu a$. The 
points in the upper plot 
correspond to $g^2\mu L = 5.66$ (diamonds), $8.84$ (pluses), 
$17.68$ (squares), and $35.36$ (x's). 
The lower plot has $g^2\mu L = 70.7$ (diamonds), $106.06$ (pluses), 
$148.49$ (squares), $212.13$ (triangles) 
and $296.98$ (x's). Lines in the lower plot are
fits of form $a-b\cdot x$. The $g^2\mu a$ ranges are different in the two 
halves. The points in the upper half are typically closer to the continuum 
limit.}
\label{eXtvsmu}
\end{figure}

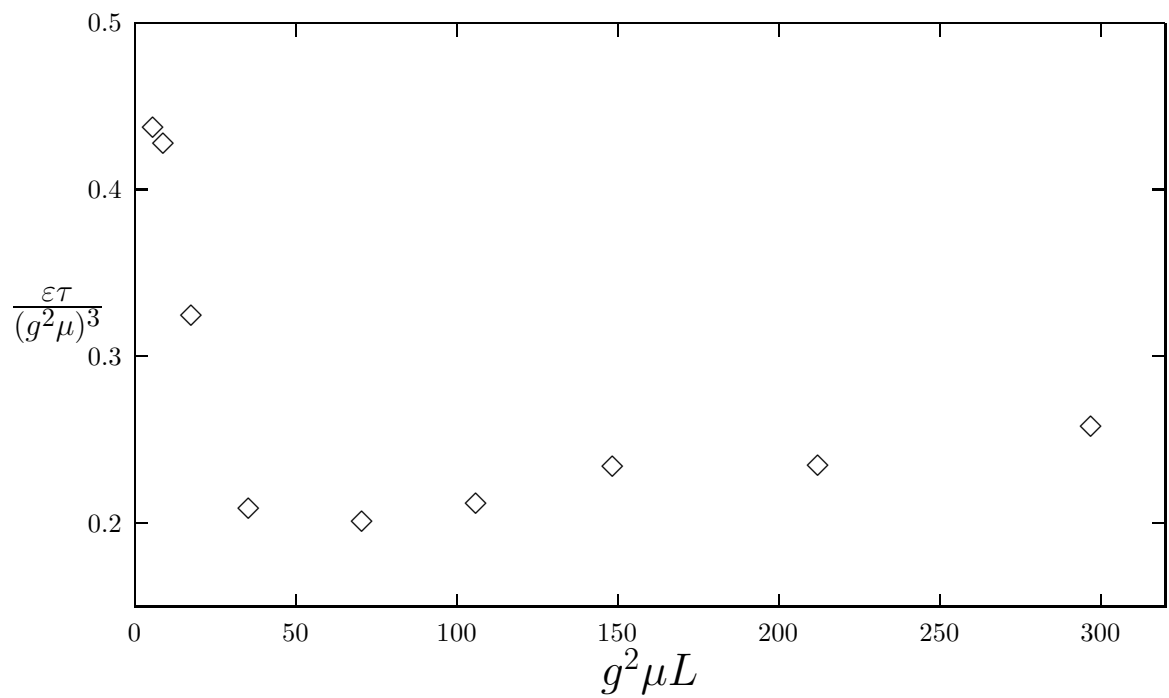
\begin{figure}[h]
\setlength{\unitlength}{0.240900pt}
\ifx\plotpoint\undefined\newsavebox{\plotpoint}\fi
\sbox{\plotpoint}{\rule[-0.200pt]{0.400pt}{0.400pt}}%
\begin{picture}(1800,1080)(0,0)
\font\gnuplot=cmr10 at 10pt
\gnuplot
\sbox{\plotpoint}{\rule[-0.200pt]{0.400pt}{0.400pt}}%
\put(161.0,254.0){\rule[-0.200pt]{4.818pt}{0.400pt}}
\put(141,254){\makebox(0,0)[r]{0.2}}
\put(1759.0,254.0){\rule[-0.200pt]{4.818pt}{0.400pt}}
\put(161.0,516.0){\rule[-0.200pt]{4.818pt}{0.400pt}}
\put(141,516){\makebox(0,0)[r]{0.3}}
\put(1759.0,516.0){\rule[-0.200pt]{4.818pt}{0.400pt}}
\put(161.0,778.0){\rule[-0.200pt]{4.818pt}{0.400pt}}
\put(141,778){\makebox(0,0)[r]{0.4}}
\put(1759.0,778.0){\rule[-0.200pt]{4.818pt}{0.400pt}}
\put(161.0,1040.0){\rule[-0.200pt]{4.818pt}{0.400pt}}
\put(141,1040){\makebox(0,0)[r]{0.5}}
\put(1759.0,1040.0){\rule[-0.200pt]{4.818pt}{0.400pt}}
\put(161.0,123.0){\rule[-0.200pt]{0.400pt}{4.818pt}}
\put(161,82){\makebox(0,0){0}}
\put(161.0,1020.0){\rule[-0.200pt]{0.400pt}{4.818pt}}
\put(414.0,123.0){\rule[-0.200pt]{0.400pt}{4.818pt}}
\put(414,82){\makebox(0,0){50}}
\put(414.0,1020.0){\rule[-0.200pt]{0.400pt}{4.818pt}}
\put(667.0,123.0){\rule[-0.200pt]{0.400pt}{4.818pt}}
\put(667,82){\makebox(0,0){100}}
\put(667.0,1020.0){\rule[-0.200pt]{0.400pt}{4.818pt}}
\put(919.0,123.0){\rule[-0.200pt]{0.400pt}{4.818pt}}
\put(919,82){\makebox(0,0){150}}
\put(919.0,1020.0){\rule[-0.200pt]{0.400pt}{4.818pt}}
\put(1172.0,123.0){\rule[-0.200pt]{0.400pt}{4.818pt}}
\put(1172,82){\makebox(0,0){200}}
\put(1172.0,1020.0){\rule[-0.200pt]{0.400pt}{4.818pt}}
\put(1425.0,123.0){\rule[-0.200pt]{0.400pt}{4.818pt}}
\put(1425,82){\makebox(0,0){250}}
\put(1425.0,1020.0){\rule[-0.200pt]{0.400pt}{4.818pt}}
\put(1678.0,123.0){\rule[-0.200pt]{0.400pt}{4.818pt}}
\put(1678,82){\makebox(0,0){300}}
\put(1678.0,1020.0){\rule[-0.200pt]{0.400pt}{4.818pt}}
\put(161.0,123.0){\rule[-0.200pt]{389.776pt}{0.400pt}}
\put(1779.0,123.0){\rule[-0.200pt]{0.400pt}{220.905pt}}
\put(161.0,1040.0){\rule[-0.200pt]{389.776pt}{0.400pt}}
\put(40,581){\makebox(0,0){\Large ${{\varepsilon\tau}\over{(g^2\mu)^3}}$}}
\put(970,21){\makebox(0,0){\Large $g^2\mu L$}}
\put(161.0,123.0){\rule[-0.200pt]{0.400pt}{220.905pt}}
\put(190,873){\raisebox{-.8pt}{\makebox(0,0){$\Diamond$}}}
\put(206,849){\raisebox{-.8pt}{\makebox(0,0){$\Diamond$}}}
\put(250,578){\raisebox{-.8pt}{\makebox(0,0){$\Diamond$}}}
\put(340,275){\raisebox{-.8pt}{\makebox(0,0){$\Diamond$}}}
\put(518,255){\raisebox{-.8pt}{\makebox(0,0){$\Diamond$}}}
\put(697,283){\raisebox{-.8pt}{\makebox(0,0){$\Diamond$}}}
\put(912,340){\raisebox{-.8pt}{\makebox(0,0){$\Diamond$}}}
\put(1234,343){\raisebox{-.8pt}{\makebox(0,0){$\Diamond$}}}
\put(1663,403){\raisebox{-.8pt}{\makebox(0,0){$\Diamond$}}}
\end{picture}
\caption{$\varepsilon\tau/(g^2\mu)^3$ extrapolated to the continuum limit: 
$f$ as a function of $g^2\mu L$. The error bars are smaller than the 
plotting symbols.}
\label{eXtvsmuL}
\end{figure}

\end{document}